\documentclass[twocolumn,showpacs,prl,amsmath,amssymb]{revtex4}

\usepackage[colorlinks,citecolor=red,linkcolor=blue,urlcolor=blue,hyperindex,bookmarks=true,bookmarksopen,pdfstartview=FitH]{hyperref}

\newcounter{loclist}
\newenvironment{qprogitemize}
 {\begin{list}{\bfseries\arabic{loclist}.}%
  {\usecounter{loclist}  
   \setlength{\labelwidth}{3em}%
   \setlength{\labelsep}{1em}%
   \setlength{\leftmargin}{3em}%
   \setlength{\parsep}{0.5ex plus 0.2ex minus 0.1ex}%
   \setlength{\itemsep}{0ex plus0.2ex}}}%
  {\end{list}}

\newenvironment{almosttrivlist}
 {\begin{list}{}%
  {\setlength{\labelwidth}{0pt}%
   \setlength{\labelsep}{0pt}%
   \setlength{\leftmargin}{2em}%
   \setlength{\parsep}{0.5ex plus 0.2ex minus 0.1ex}%
   \setlength{\itemsep}{0ex plus0.2ex}}}%
  {\end{list}}

\newcommand{\ket}[1]{|{#1} \rangle}
\newcommand{\win}[0]{\operatorname{win}}
\newcommand{\Prob}[0]{\operatorname{P}}
\newcommand{\lose}[0]{\operatorname{lose}}

\begin{document}

\title{Extended GHZ $n$-player games with 
classical probability of winning tending to 0}

\author{Michel Boyer}
\email{boyer@iro.umontreal.ca}
\affiliation{
  D\'epartement IRO, Universit\'e de Montr\'eal\\
  C.P. 6128, succursale centre-ville,\\
  Montr\'eal (Qu\'ebec) H3C 3J7 \textsc{Canada}}

\date{August 22, 2004}

\begin{abstract}
In 1990, Mermin presented a $n$ player ``game'' that is won with certainty with
$n$ spin-$\frac{1}{2}$ particles in a GHZ state, and that no classical strategy (or local theory)
can win with
probability higher than $\frac{1}{2} + \frac{1}{2^{\lceil n/2 \rceil}}$, thus establishing a new Bell
inequality.
This letter first introduces a class of arithmetic games containing Mermin's
and then gives a
quantum strategy based on a generalized $n$ party GHZ state that wins
those games with certainty.
It is then proved for a subclass of those games where each player is given
a single bit of input that no classical strategy can win
with probability higher than 
$\frac{\binom{n}{\lfloor n/2\rfloor}}{2^{n-1}}$; this is
asymptotically
$2 \sqrt{\frac{2}{\pi}} n^{-\frac{1}{2}}$,
thus giving a new and stronger Bell type inequality.
\end{abstract}



\maketitle

\paragraph{Introduction.}

In 1990, Mermin \cite{Mermin1990} proposed a ``ge\-dank\-en experiment'' with a Bell inequality \cite{Bell} that gives an exponential ratio
between expected values obtained in the quantum world and those obtainable in a local theory.
That experiment can
be viewed as a cooperative game with $n$ players that are not allowed to intercommunicate
but may share some random data:
for $j \in [1\,..\,n]$,
player $j$ is given $x_j \in \{0,1\}$ where $x = (x_j)$ is promised to be in $P$, the set
of bitstrings whose
Hamming weight $|x| = \sum_{j=1}^n x_j$ is even (in Mermin's article, the sum is odd;
what follows is \textit{mutatis mutandis}); 
player $j$ is required to output
$y_j \in \{0,1\}$; they win if the parity of $|y| = \sum_j y_j$ is that
of $|x|/2$ i.e.\ if $(-1)^{|y| + \frac{|x|}{2}} = 1$.

Again \textit{mutatis mutandis}, let the
players share the GHZ \cite{GHZ} state $\ket{\Phi_n^+} = \frac{1}{\sqrt{2}}[\ket{0^n} + \ket{1^n}]$. 
Denoting $\sigma_1$ for $\sigma_X$ and $\sigma_2$ for $\sigma_Y$ and letting
$\sigma_{1+x}$ be $\otimes_j \sigma_{1+x_j}$ we readily check that if 
$|x| \equiv 2 \pmod 4$ then
$\sigma_{1+x} \ket{\Phi_n^+} = -\ket{\Phi_n^+}$ and if $|x| \equiv 0 \pmod 4$ then
$\sigma_{1+x}\ket{\Phi_n^+} = \ket{\Phi_n^+}$. Put differently
\begin{equation}\label{opersign}
\sigma_{1+x} \ket{\Phi_n^+} = (-1)^{\frac{|x|}{2}} \ket{\Phi_n^+}
\end{equation}
This means that if they measure $\ket{\Phi_n^+}$ with $\sigma_{1+x}$ they get the parity of $\frac{|x|}{2}$:
player $j$  measures his state with $\sigma_{1+x_j}$ to get a bit $y_j$; $(-1)^{|y|} = (-1)^{\frac{|x|}{2}}$;
they win! This is the original strategy.

In order to translate it into a strategy where measurements are performed
in the standard basis  we simply notice 
that:
$\sigma_1  = H\sigma_3 H$ and
$\sigma_2  = - S^\dagger H \sigma_3 HS$ where $H$ is the Hadamard
transform and $S =\begin{bmatrix} 1 & 0 \\ 0 & i\end{bmatrix}$.
Measuring $\ket{\Phi_n^+}$ with $\sigma_{1+x}$ if $|x|$ is even is thus
measuring with $\sigma_3^{\otimes n}= \sigma_Z^{\otimes n}$ ($Z$ is the standard basis)
the state obtained 
from $\ket{\Phi_n^+}$ as follows
\begin{almosttrivlist}
\item if $x_j$ = 0, player $j$ applies $H$ on his state
\item if $x_j$ = 1, player $j$ applies  $HS$ on his state
\end{almosttrivlist}
This is the strategy as it is currently described in the quantum
information community, 
with a final step: ``player $j$ measures his state in the standard basis'' \cite{bbt03}.

In order to compare with a local model, Mermin essentially considers 
random variables $(X,Y,\Lambda)$ with $X=(X_j)$, $\Prob[X=x] = 2^{1-n}$ for $x\in P$,
$Y = (Y_j)$, the $Y_j$ being 
$\{0,1\}$-valued and independent, $Y_j$ depending  on $X_j$ and the
shared variable $\Lambda$.
Using an elegant argument, 
he shows that
$2^{n-1}\left\langle (-1)^{|Y| + |X|/2}\right\rangle \leq
2^{\lfloor n/2\rfloor}$; i.e.
\begin{equation}\label{merminexpbound}
  \left\langle (-1)^{|Y|+|X|/2}\right\rangle \leq 
  2^{\lfloor n/2\rfloor -n +1} = \frac{1}{2^{\lceil n/2\rceil-1}}
\end{equation}
Compared to  1 with the quantum strategy, this bound gives an exponential ratio.
However,
the random variable
$(-1)^{|Y|+|X|/2}$ takes only two values, $1$ and $-1$, $1$ corresponding to
``win'' and $-1$ to ``lose'' and so
\[
  \left\langle (-1)^{|Y|+|X|/2}\right\rangle = \Prob[\win] - \Prob[\lose] =
   2\Prob[\win] -1
\]
and (\ref{merminexpbound}) is equivalent to
\begin{equation}\label{merminloosebound}
\Prob[\win] \leq \frac{1}{2} + \frac{1}{2^{\lceil n/2\rceil}}
\end{equation}
a result that was subsequently derived independently using
combinatorial methods in \cite{bbt03} which shows the
bound is tight (\ref{mermintightbound}). 
This means that in terms of probabilities, the quantum strategy wins with
a probability that is not even twice that of a well chosen
classical strategy! This letter is meant to widen that gap,
introducing ``arithmetic games'' that contain Mermin's game, as well as an extension
presented in \cite{BHMR}, and that also contain extensions in an other
direction that behave very nicely and for which
the probability of winning with classical strategies tends to 0 as $n\to \infty$.

\paragraph{A n-player game modulo $M$ with divisor $D$.}
Each of $n\geq 1$ players is given as input
an integer $x_j \in [0\,..\,D-1]$ for some 
fixed integer $D\geq 2$ (the ``divisor''); 
they are promised that $D$ divides their sum, i.e.
\begin{equation}\label{promise}
\sum_{j=1}^n x_j \equiv 0 \pmod D
\end{equation}
The answers the players are allowed to give 
are integers $y_j\in [0\,..\,M-1]$ for some other fixed integer $M\geq 2$
(the ``modulo'').
The game is won if their answers
satisfy
\begin{equation}\label{winningcondition}
\sum_{j=1}^n y_j \equiv \frac{\sum_j x_j}{D} \pmod M
\end{equation}
We denote this game $G_{n,D,M}$. Game $G_{n,2,2}$ is
Mermin's $n$-player 
game.
Games $G_{n,2^d,2}$ (divisor $2^d$ and modulo 2)
were presented in \cite{BHMR}.
We will be interested by the games $G_{n,2,2^m}$, i.e.\ with
divisor $D=2$ and modulo $M=2^m$ but our quantum
strategy works for any $D$ and $M$. 
Let $P$ be the set of the $x$ in $[0\,..\,D-1]^n$ that satisfy (\ref{promise})
and $W$ be the set of the $(x,y)$ with $x\in P$ and $y\in Y = [0\,..\,M-1]^n$
that satisfy (\ref{winningcondition});
let finally $W(x) = \{y \in Y \mid (x,y) \in W\}$ for 
$x\in P$: these are the outputs that win on input $x$.

\paragraph{The quantum strategy.}

For any integer $M$, we let
$\omega_M = e^{2i\pi/M}$, the standard $M$-th primitive root of the unity.
Knowing the game in advance, the
$n$ players have prepared the following generalized GHZ state 
\[
\ket{\Phi} = \frac{1}{\sqrt{M}} \sum_{k=0}^{M-1} \,\ket{k}^{\otimes n}
\]
which they now share. For
$0 \leq k < M$ and $M'$ arbitrary (to be chosen appropriately later on)
let
\begin{align*}
 F_{M}\ket{k} &= \frac{1}{\sqrt{M}} \sum_{y=0}^{M-1} \omega_{M}^{ky} \,\ket{y} \\
 S_{M'}\ket{k} &= \omega_{M'}^k \,\ket{k}
\end{align*}
where $ky$ is the integer multiplication of $k$ by $y$. $F_{M}$ is the quantum
Fourier transform \cite{NielsenChuang} on $\mathbf{C}^{[0\,..\,M-1]}$. 
For instance, if $D=2$ and $M=2$,  $F_M^\dagger = H$
and $S_{DM} = S_4 
= S$.
Now here is the strategy. Notice that for $D=2$ and $M=2$, it is
exactly
that of Mermin's game as recalled in the introduction
($H = F_2^\dagger (S_4)^0 $, $HS = F_2^\dagger (S_4)^1$).
\begin{qprogitemize}
\item Player $j$ applies $(S_{DM})^{x_j}$ to his  state
\item Player $j$ applies $F^\dagger_{M}$ to his state
\item Player $j$ measures his state in the standard basis $\ket{0},\ldots,\ket{M-1}$ to output $y_j$.
\end{qprogitemize}
Given initial state $\ket{\Phi}$, 
using $(\omega_{DM})^D = \omega_{M}$,
one checks immediately that the states after
steps \textbf{1} and \textbf{2} are respectively:
\begin{qprogitemize}
\item
  $\frac{1}{\sqrt{M}} \sum_{k=0}^{M-1} 
   \omega_{M}^{k\frac{\sum_{j=1}^n x_j}{D}} \ket{k}^{\otimes n}$ \\
\item $\kappa \sum_{k,y_1\ldots y_n} 
   \omega_{M}^{k\frac{\sum_{j=1}^n x_j}{D}} \omega_{M}^{-k \sum_{j=1}^n y_j} \ket{y_1\ldots y_n}$ \hfill\refstepcounter{equation}\label{bigsum}(\theequation)
\end{qprogitemize}
with normalising factor $\kappa = (\sqrt{M})^{-(n+1)}$.
The coefficient of 
$\ket{y} = \ket{y_1,\ldots, y_n}$ in (\ref{bigsum}) is
\[
 \kappa \sum_{k =0}^{M-1}
   \omega_{M}^{k\left(\frac{\sum_{j=1}^n x_j}{D}-\sum_{j=1}^n y_j\right)}
\]
and it is
$0$ unless $\sum_{j=1}^n y_j \equiv \frac{\sum_{j=1}^n}{D} \pmod M$, i.e.\ unless
$y \in W(x)$, in which case it is
$\kappa M$ and all the coefficients of $\ket{y}$ for $y\in W(x)$ are then equal;
state (\ref{bigsum}) is thus
\[
  \frac{1}{|W(x)|^{\frac{1}{2}}}\,\sum_{y \in W(x)} \,\ket{y}
\]
(because the $\ket{y}$ are orthonormal).
When the players measure in step \textbf{3}, they get with certainty $\ket{y}$ such that
$y\in W(x)$ and collectively win the game without any need to intercommunicate.

\paragraph{Quantum pseudo-telepathy games.} 
Our notations come from
\cite{bbt03,bbt04}.
Those are $n$ player cooperative games $G$
where intercommunication is disallowed and such that: (1) there
is a quantum strategy that wins with certainty, (2) there is no classical
strategy that wins with certainty. 
A classical (i.e.\ non quantum) strategy $\mathcal{S}$ is \textit{deterministic}
if each player $j$ returns a predetermined output $\mathcal{S}_j(x_j)$ on input $x_j$ \footnote{For $G_{n,D,M}$ games
a deterministic strategy $\mathcal{S}$ is identified to $n$ functions $\mathcal{S}_j: [0\,..\,D-1] \to [0\,..\,M-1]$ for $j\in [1\,..\,n]$.}; 
we denote $\widetilde{\omega}(G)$ the maximum rate of success of all the deterministic strategies;
letting
$\mathcal{S}(x) = (\mathcal{S}_j(x_j))$, it is
\[
\widetilde{\omega}(G) = \max_{\mathcal S} \frac{| \{x \in P \mid \mathcal{S}(x) \in W(x)\}|}{|P|}
\]
A strategy is \textit{probabilistic} if the players may also access a shared random value $s$ 
and $\mathcal{S}(x,s) = (\mathcal{S}_j(x_j,s)) $\footnote{We include in $s$
the random generators (if any) 
of all players so that $\mathcal{S}^s$ defined by $\mathcal{S}^s(x) = \mathcal{S}(x,s)$ is deterministic for all
$s$.}; we write $\Prob_{\mathcal{S}}[\win\mid x]$
for $\Prob[\{s \mid \mathcal{S}(x,s) \in W(x)\}]$ and define
\begin{equation}\label{defomega}
\omega(G) = \max_{\mathcal S} \min_{x\in P} \Prob_{\mathcal S}[\win\mid x]
\end{equation}
One shows easily that $\omega(G)\leq \widetilde{\omega}(G)$ \cite{bbt04}. A game is a pseudo-telepathy game
if and only if it has a quantum winning strategy and $\widetilde{\omega}(G) < 1$. 
Bound (\ref{merminloosebound}) and its tightness
can now be stated precisely \cite{Broadbent} \footnote{The proof that $\omega(G_{n,2,2}) =
\widetilde{\omega}(G_{n,2,2})$ is not trivial.}: if $n \geq 1$
\begin{equation}\label{mermintightbound}
\omega(G_{n,2,2}) = \widetilde{\omega}(G_{n,2,2}) = \frac{1}{2} + \frac{1}{2^{\lceil n/2\rceil}}
\end{equation}

\paragraph{The $G_{n,D,M}$ games and pseudo-telepathy.}
First, the game $G_{n,D,M}$ is not a pseudo-telepathy game if $\gcd(D,M) = 1$.
Indeed
Bezout's theorem implies that there is then $a\in [0\,..\,M-1]$ such that
and $aD \equiv 1 \pmod M$.
If player $j$ answers $\mathcal{S}_j(x_j) = ax_j \mod M$ where
$x \mod M$ stands for  $x - M\lfloor x/M\rfloor$  then for $x\in P$
\[
\sum_j \mathcal{S}_j(x_j) \equiv a\sum_j x_j \equiv aD \frac{\sum_j x_j}{D} \equiv \frac{\sum_j x_j}{D} \pmod M
\]
and ${\mathcal S}(x) = (\mathcal{S}_j(x_j))$ wins the game on all valid inputs.

If $\gcd(D,M) \neq 1$ then $G_{n,D,M}$ is a pseudo-telepathy game if
$n \geq \max(3,p)$ where $p$ is the smallest common prime factor of $M$ and $D$.
Let indeed $D = pD'$ and $M = pM'$ and
let us assume $\mathcal{S}$ is a deterministic
winning strategy for $G_{n,D,M}$; we want to derive a contradiction.
Let $\mathcal{S}'_j(x_j) = \mathcal{S}_j(D'x_j)\mod p$.
If $\sum_j x_j \equiv 0 \pmod p$ then $\sum_j D'x_j \equiv 0 \pmod D$ and
by assumption
\[
\sum_j \mathcal{S}_j(D'x_j) \equiv \frac{\sum_j D'x_j}{D} \equiv \frac{\sum_j x_j}{p} \pmod {pM'}
\]
and, since $x \equiv y \pmod {pM'}$ implies $x \equiv y \pmod {p}$
\[
\sum_j \mathcal{S}'_j(x_j) \equiv \frac{\sum_j x_j}{p} \pmod p
\]
and $\mathcal{S}'$ is then a winning strategy for $G_{n,p,p}$. 
Let now $j_1,j_2\in [1\,..\,n]$ be any two players; then
$\mathcal{S}'_{j_1}(1) - \mathcal{S}'_{j_1}(0) \equiv
\mathcal{S}'_{j_2}(1) - \mathcal{S}'_{j_2}(0) \pmod p$; we can see this by giving
all the other players inputs so that the overall total is $p$ (we need $n\geq 3$).
This means there is some $d \in [0\,..\,p-1]$ such that 
$\mathcal{S}'_{j}(1) - \mathcal{S}'_{j}(0) \equiv d \pmod p$ for all players $j$.
On the other hand if we choose $x_j = 1$ for $1\leq j \leq p$ and 0 otherwise, their sum is $p$ and
$\sum_{j=1}^p \mathcal{S}'_j(1) + \sum_{j=p+1}^n \mathcal{S}'_j(0) \equiv 1 \pmod p$.
If we choose all $x_j = 0$ their sum is 0 and
$\sum_{j=1}^n \mathcal{S}'_j(0) \equiv 0 \pmod p$. Subtracting those two identities leaves
\[
\sum_{j=1}^p \left[ \mathcal{S}'_j(1) - \mathcal{S}'_j(0)\right] \equiv 1 \pmod p
\]
i.e.\ $pd \equiv 1 \pmod p$. This implies that $p$ divides 1 and gives the desired contradiction.
Written differently, $\widetilde{\omega}(G_{n,D,M}) < 1$ if $n \geq \max(3,p)$. 

We now
select a subclass of those games that give a better bound than (\ref{mermintightbound}).
\paragraph{A bound on classical strategies.}
From now on we consider only the games $G_{n,2,2^m}$ with $m\geq 1$; the divisor $D$ is
2, the  modulo is a power of 2.
The inputs $x\in [0\,..\,D-1]^n$ are consequently bitstrings of length $n$
and the promise is that $|x| = \sum_{j=1}^n x_j$ is even; 
$|P|$, the number of possible inputs, is thus $2^{n-1}$.
We now show that if $\lceil n/2\rceil < 2M$ with $M=2^m$ then any classical strategy
wins on at most
$\binom{n}{\lfloor n/2\rfloor}$ inputs or, equivalently, for $n \geq 1$
\begin{align}\label{classicalbound}
\widetilde{\omega}(G_{n,2,2^m}) &\leq \frac{\binom{n}{\lfloor n/2\rfloor}}{2^{n-1}} &\text{if $m \geq \ell(n)$}
\end{align}
where $\ell(n) = \lfloor \lg((n+1)/4) \rfloor + 1$.
Let indeed $\mathcal{S}$ be any
fixed deterministic strategy; since $x_j$ is either 0 or 1
\[
\sum_{j=1}^n \mathcal{S}_j(x_j) = \sum_{j=1}^n \left[ \mathcal{S}_j(0)(1-x_j) + \mathcal{S}_j(1)x_j\right]
\]
If we let $d_j = \mathcal{S}_j(1) - \mathcal{S}_j(0)$ and $b = \sum_{j=1}^n \mathcal{S}_j(0)$, we 
get
\[
\sum_{j=1}^n \mathcal{S}_j(x_j) = \sum_{j=1}^n d_j x_j + b
\]
Strategy $\mathcal{S}$ wins on input $x$ if and only if
 $\sum_{j=1}^n d_jx_j + b \equiv \frac{\sum_{j=1}^n x_j}{2} \pmod M$
or, equivalently, if and only if
\begin{equation}\label{winone}
 \sum_{j=1}^n (2d_j - 1) x_j \equiv -2b \pmod {2M}
\end{equation}
We now use a theorem of Griggs \cite{griggs}: 
\textit{Let $q>0$ be any
integer and $a_1$,\ldots, $a_n$ be integers that have no common factor with $q$
(i.e.\ $\gcd(a_j,q)=1$). Let $E$ be any subset of $r$ elements of
$[0\,..\, q-1]$. Then the number of strings $x\in\{0,1\}^n$ such that
$\sum_{j=1}^n a_jx_j$ is congruent modulo $q$ to some element  in $E$ is at most
$\sum_{j=\lceil (n-r)/2\rceil}^{\lceil (n+r)/2\rceil-1} \binom{n}{j}_q$
where $\binom{n}{j}_q$ is the number of bitstrings $s\in \{0,1\}^n$ such
that $|s| \equiv j \pmod q$,
and this bound is tight.}

Letting $q$ be $2M$ and 
$a_j = 2d_j-1$, $a_j$ is odd and consequently relatively prime with $q = 2M = 2^{m+1}$.
The theorem applies to (\ref{winone}) and
the number of winning inputs ($r$ being 1) 
is at most $\binom{n}{\lfloor n/2\rfloor}_{2M}$
which is equal to $\binom{n}{\lfloor n/2\rfloor}$
as soon as $\lceil n/2\rceil < 2M$ i.e.\ $m\geq \ell(n)$. This proves our claim \footnote{If 
$M$ were not a power of 2, there would be no way to prevent the players from choosing
$a_j$ that are factors of $2M$. For instance, if $M$ is $22$, and the players choose
$d_j=6$ and so $a_j=11$, then $\sum_j a_jx_j \equiv 11|x| \pmod {44}$ which is $0 \pmod {44}$ if $|x|/2$ is
even and $22 \pmod {44}$ if $|x|/2$ is odd, and there results a
strategy with more than $50\%$ of winning inputs: player 1 answers 11 if $x_1=0$ and 17 otherwise
whilst the other players answer 0 if $x_j=0$ and 6 otherwise. 
The same reasoning applies
for every $M = 2(2r+1)$, $r \geq 1$, giving $50\%$ or more of winning inputs.}.
From now on $\lceil n/2\rceil < 2M$ with $M=2^m$.

\paragraph{An optimal probabilistic strategy.}

The game is to
halve $|x|$ i.e.\ the number of ones in the bitstring $x$, the players giving their
answers using $m$ bit integers. With unary representation, it makes sense
to halve using random methods.  This is the intuition behind the following strategy.
Let
\[
S = \{s \in \{0,1\}^n \text{\ \ s.t.\ $|s|$ has the same parity as $\lfloor n/2\rfloor$}\}
\]
where $|s| = \sum_{j=1}^n s_j$;  $|S| = 2^{n-1}$. 
We now define the probabilistic strategy
$\mathcal{S}$ as follows: 
the players draw randomly (with equal probability) $s\in S$
(in fact, they have drawn beforehand and just look up what they have 
drawn \footnote{Each player needs only keep 1 bit $s_j$ for each round of the game
except for player 1 who needs both $s_1$ and $|s|$.
Choosing $s$ implies no communication (after the game is started).})
and then use the deterministic strategy $\mathcal{S}^s$
where $\mathcal{S}^s(x) = \mathcal{S}(x,s) = (\mathcal{S}_j(x_j,s))$ is such that:
\begin{almosttrivlist}
\item Player $j$ answers $\mathcal{S}_j(x_j,s) = s_jx_j$ for $2 \leq j \leq n$
\item Player $1$ answers $\mathcal{S}_1(x_1,s) = s_1x_1 + (\lfloor n/2\rfloor - |s|)/2 \mod M$
\end{almosttrivlist}

\paragraph{The winning condition.}
By definition, $\mathcal{S}^s(x)\in W(x)$ %
($\mathcal{S}^s$ wins on input $x$) if and only if
\[
\sum_{j=1}^n s_j x_j  + (\lfloor n/2\rfloor - |s|)/2 \equiv |x|/2 \pmod M
\]
or equivalently
$
|s| + |x| - 2\sum_j s_jx_j \equiv \lfloor n/2\rfloor \pmod {2M}
$
which holds if and only if
\begin{equation}\label{stringwinningcondition}
 | s\oplus x|  = \lfloor n/2\rfloor 
\end{equation}
where $s \oplus x$ is the exclusive or of the bitstrings $s$ and $x$ 
(or equivalently their sum in $(\mathbf{Z}/2\mathbf{Z})^n$) 
and strict equality must occur because 
$0\leq |s \oplus x| \leq n$ and $\lceil n/2\rceil < 2M$ \footnote{The 
distance between $|s\oplus x|$ and $\lfloor n/2\rfloor$ is always less than $2M$:
$|s\oplus x| - \lfloor n/2\rfloor \leq n - \lfloor n/2\rfloor \leq
\lceil n/2\rceil < 2M$ and $|0 - \lfloor n/2\rfloor| = \lfloor n/2\rfloor < 2M$. 
}.

Let $\Omega$ be the set of strings $w\in \{0,1\}^n$ such that
 $|w| = \lfloor n/2\rfloor$; $|\Omega| = \binom{n}{\lfloor n/2\rfloor}$
and by (\ref{stringwinningcondition})
for all $x\in P$, $s \in  S$
\begin{equation}\label{duality}
 \mathcal{S}^s(x) \in W(x) \iff x \oplus s \in \Omega
\end{equation}
\paragraph{Proportion of winning inputs.}
For any $s\in S$, the proportion of winning inputs $x$ for $\mathcal{S}^s$ is the number of
those $x\in P$ such that $\mathcal{S}^s(x) \in W(x)$
divided by $|P|$, and by (\ref{duality}),
since $x\oplus s = w$ iff $x = s\oplus w$, it is
\[
\frac{|\{s \oplus w \mid w \in \Omega\}|}{|P|} = \frac{\binom{n}{\lfloor n/2\rfloor}}{2^{n-1}}
\]
According to (\ref{classicalbound}) this is the maximum possible and so
\begin{align*}
\widetilde{\omega}(G_{n,2,2^m}) &= \frac{\binom{n}{\lfloor n/2\rfloor}}{2^{n-1}} &\text{if $n\geq 1$ and $m \geq \ell(n)$}
\end{align*}
\paragraph{Probability of winning on input $x$.}
For any $x\in P$, $\Prob_{\mathcal{S}}[\win\mid x]$ i.e.\ $\Prob[\{s\mid \mathcal{S}(x,s)\in W(x)\}]$
is simply the 
proportion of the strategy strings $s\in S$ such that $\mathcal{S}(x,s) \in W(x)$.
Using again (\ref{duality}) this probability is
\begin{equation}\label{optimalproportion}
 \Prob_{\mathcal{S}}[\win \mid x] = 
 \frac{|\{x\oplus w \mid w\in \Omega\}|}{|S|} = 
\frac{\binom{n}{\lfloor n/2\rfloor}}{2^{n-1}}
\end{equation}
which must then also be $\min_{x\in P} \Prob_{\mathcal{S}}[\win\mid x]$.
By (\ref{defomega}) this gives a lower bound for $\omega(G_{n,2,2^m})$; by 
 $\omega(G)\leq \widetilde{\omega}(G)$ and (\ref{classicalbound}) it is also an
upper bound and so
for $n \geq 1$, $m \geq \ell(n)$
\begin{equation}\label{boyertightbound}
\omega(G_{n,2,2^m}) = \widetilde{\omega}(G_{n,2,2^m}) = \frac{\binom{n}{\lfloor n/2\rfloor}}{2^{n-1}}
\end{equation}
\paragraph{An asymptotic bound.}
Using Stirling formula  $n! \sim (2\pi)^{\frac{1}{2}}n^{n+\frac{1}{2}} e^{-n}$
we obtain for $n=2k$
\[
\frac{\binom{2k}{k}}{2^{2k-1}} = \frac{(2k)!}{2^{2k-1}k!k!} \sim 2 (\pi k)^{-\frac{1}{2}}
\]
For $n = 2k+1$ we have $\binom{2k+1}{k} = 
\frac{2k+1}{k+1}\binom{2k}{k}$ and thus ${\displaystyle \lim_{k \to \infty}} \frac{\binom{2k+1}{k}/2^{2k}}{\binom{2k}{k}/2^{2k-1}} = 1$ 
which
implies that 
\begin{equation}\label{boyerasymptoticbound}
  \omega(G_{n,2,2^{\ell(n)}}) = \widetilde{\omega}(G_{n,2,2^{\ell(n)}}) 
  \sim 2 \sqrt{\frac{2}{\pi}} n^{-\frac{1}{2}}
\end{equation}
with $\ell(n) = \lfloor \lg((n+1)/4) \rfloor + 1$,
the minimum  length of the binary numbers
to be used as answers by the players.

\paragraph{Conclusion.}
First, we have extended the class of known pseudo-telepathy games.
Then, considering the class of $n$ player games modulo $2^m$ and divisor $2$ with
$n \geq 1$ and $m \geq \ell(n)$, we have much improved 
the proven gap (\ref{mermintightbound}) between what can be
done quantumly and classically: the optimal probability (\ref{boyertightbound}) that $n$ players win
our one bit of input games with classical means tends to 0 as $n\to \infty$. 
We exhibited
a quantum strategy that works for any divisor $D$ and any modulo $M$. The 
simplicity of our classical analysis for $D=2$ and $M=2^m$ may depend
on the simplicity of bitstrings; it is as yet unclear how our method
could be extended to trits ($D=3$, $M=3^m$).

On a practical basis, the bound depends only on the number of players as soon
as the surprisingly weak condition $m \geq \ell(n)$ or equivalently $\lceil n/2\rceil < 2^{m+1}$ is met; this means that we essentially win at least 
one bit over what we would naturally
expect. Indeed, with a 2 bit implementation, i.e.\ $m=2$, we get $\lceil n/2\rceil < 2^3$ iff $n \leq 14$;
with $n=14$, the integer ``halves'' go from 0 to 7 and not from 0 to 3, but as we have seen not only in the proof of
the bound but also in the independent proof of (\ref{stringwinningcondition}) and (\ref{duality}), residues modulo 4 give the same bound.
This gain may be useful: not only do we need $m$ bit answers, but
we also need to implement the quantum strategy, and in particular the Fourier transforms, on
$m$ qubits. We need $n=7$ ($m=2$)
to beat Mermin's bound, $n=9$ ($m=2$) to go below $50\%$ and $n=41$ ($m=4$) to go below $25\%$
of winning inputs.

\begin{acknowledgments}
Insightful comments from Anne Broadbent are gratefully acknowledged.
\end{acknowledgments}

\end{document}